\documentclass[12pt]{iopart}
\pdfoutput=1

\usepackage{graphicx, epsfig,epstopdf}
\usepackage{bm}
\usepackage{mathrsfs}
\usepackage{ulem}

\newcommand{\ket}[1]{\left\vert#1\right\rangle}
\newcommand{\bra}[1]{\left\langle#1\right\vert}
\newcommand{\eqref}[1]{(\ref{#1})}

\begin{document}
\title[Dynamics of the entanglement spectrum in spin chains]{Dynamics of the entanglement spectrum in spin chains}
\author{G. Torlai$^1$}
\address{$^1$Department of Physics and Arnold Sommerfeld Center for Theoretical Physics,  Ludwig Maximilian Universit\"at M\"unchen, Theresienstr. 37, 80333 Munich, Germany.}
\author{L. Tagliacozzo$^2$}
\address{$^2$ ICFO, Av. Carl Friedrich Gauss 3, 08860 Castelldefels (Barcelona), Spain.}
\author{G. De Chiara$^3$}
\address{$^3$ Centre for Theoretical Atomic, Molecular and Optical Physics,
School of Mathematics and Physics, QueenÕs University, Belfast BT7 1NN, United Kingdom}
\ead{g.dechiara@qub.ac.uk}
\begin{abstract}
We study the dynamics of the entanglement spectrum, that is the time evolution of the eigenvalues of the reduced density matrices after a bipartition of a one-dimensional spin chain. Starting from the ground state of an initial Hamiltonian, the state of the system is evolved in time with a new  Hamiltonian. We consider both instantaneous and quasi adiabatic quenches of the system Hamiltonian across a quantum phase transition. We analyse the Ising model that can be exactly solved and the XXZ for which we employ the time-dependent density matrix renormalisation group algorithm. Our results show once more a connection between the Schmidt gap, i.e. the difference of the two largest eigenvalues of the reduced density matrix, and order parameters, in this case the spontaneous magnetisation.
\end{abstract}



\maketitle

\section{Introduction}
In the last decade, entanglement is revealing itself an extremely useful tool in the characterisation of quantum many-body systems \cite{AmicoRMP}. Since the full-tomography of a quantum state would require an exponentially large number of measurements, measures of quantum correlations, including entanglement, give a lot more information than expectation values of local operators. Many works have concentrated on bipartite entanglement including von Neumann and Renyi entropies \cite{CalabreseReview,Coser}, mutual information \cite{mutual_info}, negativity between blocks of particles \cite{Calabrese2012}, concurrence between two spin-1/2 particles as first studied in \cite{Osborne2002,Osterloh2002}.
Beyond entanglement, the study of discord and its global multipartite extension in spin chains have also received recently a lot of attention \cite{varie,Rulli,Campbell}. The multipartite setting is far less explored because of its complexity compared to the bipartite scenario, although a few works show the emergence of genuine multipartite entanglement close to quantum phase transitions in spin chains \cite{GME,Julia}.

In the last few years, entanglement spectrum (ES), consisting of the eigenvalues of the reduced density matrix of one of the two parties in a bipartition of the Hilbert space, for example in real or momentum space has received enormous interest in the condensed matter physics community. For strongly correlated systems, ES was studied in relation to the performance of the density matrix renormalisation group~\cite{white93}. Since Li and Haldane's seminal paper \cite{Li08},  many authors have investigated the ES in a variety of settings including one-dimensional spin chains \cite{ES_bib}. In particular, in critical 1D systems described by a conformal field theory, the density distribution of the eigenvalues of the reduced density matrix is universal and only depends on the central charge of the corresponding conformal theory \cite{Lefevre}.  However, it is important to stress that the discrete part of the ES, containing the largest eigenvalues, does not follow this universal distribution, and thus cannot be predicted using the von Neumann entropy. Instead this region of the ES can be connected to the energy spectrum of a boundary conformal field theory \cite{Lauchli2013}. Other studies have shown that the analysis of the individual eigenvalues gives much more information than it was anticipated. In particular the so-called Schmidt gap, the difference of the two largest eigenvalues, detects very accurately the position of a critical point and scales with universal critical exponents \cite{DeChiara2012,Lepori2013,Lundgren,Santos,Giampaolo2013, Anna2013}. The Schmidt gap can also be employed in numerical simulations to reveal Haldane phases that lack a local order parameter. As in these hidden order phases the ES must be formed of evenly degenerate multiplets \cite{Pollmann10}, the Schmidt gap is bound to close \cite{DeChiara11}.
Since this information is not accessible in the entanglement entropy, the study of the ES has deepen the understanding of many-body quantum states.

The dynamics of strongly correlated systems has also received particular attention recently. On one hand, one question is whether or not isolated quantum systems reach a steady state after a sudden change of the Hamiltonian (for a review check \cite{PolkoRMP}) and whether this state resembles a thermal state. Ref.~\cite{fagotti2013} studied the evolution of the reduced density matrix  in the quantum Ising model. On the other hand, quasi adiabatic sweeps of a Hamiltonian parameter across the transition reveals universal dynamics ruled by the Kibble-Zurek mechanism (see \cite{Adolfo2013} for recent reviews). 

In this work we analyse the time evolution of the ES in a system of 1D spin chains after instantaneous or quasi-adiabatic quenches in one parameter of the Hamiltonian. We first study the 1D Ising model that can be exactly diagonalised and we show results on the evolution of the ES after a sudden change of the magnetic field. We show that when the magnetic field is changed from the paramagnetic phase to the ferromagnetic phase, the ES levels cross each other and the corresponding Schmidt gap oscillates. The oscillations are equivalent to the oscillations of the spontaneous magnetisation after a similar quench. We also analyse the evolution of ES after a quasi-adiabatic quench and show the appearance of an algebraic dependence of the Schmidt gap with the rate of change of the magnetic field when crossing the quantum phase transition.

We then move to the XXZ model in which we quench the value of the anisotropy parameter. The unitary evolution of the ground state in this case is calculated using the time-dependent density matrix renormalisation group (tDMRG) algorithm \cite{dmrg}. In this case however we do not observe crossing of the eigenvalues but rather a tendency of them to become degenerate and at the same time decreasing exponentially with time.

The paper is organised as follows: in Sec.~\ref{sec:prel} we present the models we consider in this work and review the basic properties of the ES; in Sec.~\ref{sec:Ising} we present our results for the dynamics of the Ising model; in Sec.~\ref{sec:XXZ} we pass to the ES of the XXZ model and in Sec.~\ref{sec:conc} we summarise and conclude. In the Appendix, details of the diagonalisation of the Ising model can be found.

\section{Preliminaries}
\label{sec:prel}
In this paper, we will study the Hamiltonian of two spin chain models. The first is the spin-1/2 transverse field Ising model described by the quantum Hamiltonian:
\begin{equation}
\label{ising}
\hat{H}(h)=-\frac{1}{2}\Bigl[\sum_{i=1}^{L-1}\,\hat{\sigma}_i^x\hat{\sigma}_{i+1}^x+h\sum_{i=1}^L\,\hat{\sigma}_i^z\Bigl]\qquad\qquad h>0.
\end{equation}
with $L$ the number of spins in the chain and $h$ proportional to the magnetic field oriented in the $z$ direction and throughout this paper we always assume open boundary conditions. Operators $\hat{\sigma}_i^{x,y,z}$ are the Pauli spin operators of site $i$.
As the parameter $h$ approaches 1 (quantum critical point), the system undergoes a quantum phase transition between a ferromagnetic phase ($h<1$) and a quantum paramagnetic phase ($h>1$). In the paramagnetic phase the ground state is $Z_2$ invariant, and thus completely disordered, while for the ferromagnetic phase, in the thermodynamic limit, the ground state breaks $Z_2$ invariance (see for example \cite{Sachdev}). This model is integrable and can be exactly diagonalised using the well known Jordan-Wigner and Bogoliubov transformations, as described in  ~\ref{sec:ff}.

The second model is the spin-1/2 XXZ model of a 1D chain with Hamiltonian:
\begin{equation}
\label{eq:XXZ}
\hat{H}(\Delta)=\sum_{i=1}^{L-1}\,\left[\hat{\sigma}_i^x\hat{\sigma}_{i+1}^x+\hat{\sigma}_i^y\hat{\sigma}_{i+1}^y+\Delta\hat{\sigma}_i^z\hat{\sigma}_{i+1}^z\right]
\end{equation}
and anisotropy $\Delta$. The phase diagram of this model is very well known: the ground state of Hamiltonian (\ref{eq:XXZ}) is ferromagnetic for $\Delta<-1$ and N\'eel antiferromagnetic for $\Delta>1$. In the intermediate region $-1<\Delta<1$ the system is in the critical-XY phase, characterised by absence of energy gap, divergence of the correlation length, logarithmic growth of the block entanglement entropy with the size of the block.

We assume the system to be initially at time $t=0$ in the ground state $\ket{\Psi_0}$ of $\hat{H}_0$ that is Hamiltonian Eq.~(\ref{ising}) with magnetic field $h_0$ or Hamiltonian Eq.~(\ref{eq:XXZ}) with initial anisotropy $\Delta_0$. The system is then evolved with a quench in the magnetic field or the anisotropy, respectively. We consider two scenarios. In the first one, the magnetic field $h$ (anisotropy $\Delta$) is instantaneously changed to a final value $h_1$ ($\Delta_1$) and the system is let to evolve with the new Hamiltonian $\hat{H}_1$ that is Hamiltonian Eq.~(\ref{ising}) with magnetic field $h_1$ or Hamiltonian Eq.~(\ref{eq:XXZ}) with final anisotropy $\Delta_1$. In other words the state of the system at time $t$ is:
\begin{equation}
\ket{\Psi(t)} = e^{-i \hat H_1 t} \ket{\Psi_0}
\end{equation}
 and we assume $\hbar=1$ for simplicity.

In the second scenario, and only for the Ising model, we assume that the magnetic field changes linearly in time with rate $\gamma$: 
\begin{equation}
\label{eq:htgamma}
h(t) = h_0-\gamma t
\end{equation}
until the final time $T=(h_0-h_1)/\gamma$. Notice that $\gamma>0$ when $h_0>h_1$ and viceversa. In this case the state of the system follows Schr\"odinger equation:
\begin{equation}
i\frac{d}{dt}\ket{\Psi(t)} =\hat H(h(t))  \ket{\Psi(t)}.
\end{equation}

The entanglement spectrum of a state $\ket \Psi$ is obtained by tracing out half of the chain and computing the eigenvalues of the reduced density matrix of one of the two halfs. In other words, we partition the chain in two parts $\Omega$ and $\Omega^\perp$ containing the spins $1,2,\dots, L/2$ and $L/2+1,\dots, L$, respectively, assuming for simplicity even $L$. 
Then we compute the eigenvalues of the matrix:
\begin{equation}
\rho_\Omega ={\rm Tr}_{\Omega^\perp} \ket{\Psi}\bra{\Psi}.
\end{equation}
As ${\rm Tr}\rho_\Omega=1$, the sum of the eigenvalues $\lambda_1,\lambda_2,\dots$, sorted in decreasing order, is equal to 1.
As shown in Ref.~\cite{DeChiara2012}, for a large homogeneous chain, the exact position of the cut, as long as it is far from the chain edges, does not affect the results.
In the next sections we will analyse the two models separately.

\section{Quenches in the Ising model}
\label{sec:Ising}
In this section we discuss the results for the quenches in the Ising Hamiltonian. The details of the calculation can be found in the Appendix. We first consider instantaneous quenches and then move to quasi-adiabatic quenches.

\subsection{Results for the Ising model: instantaneous quenches}

In this section we discuss the results obtained for the ES after an instantaneous quench $h_0\to h_1$ for $L=200$. 
\begin{figure}[h]
\centering
\includegraphics[width=145mm]{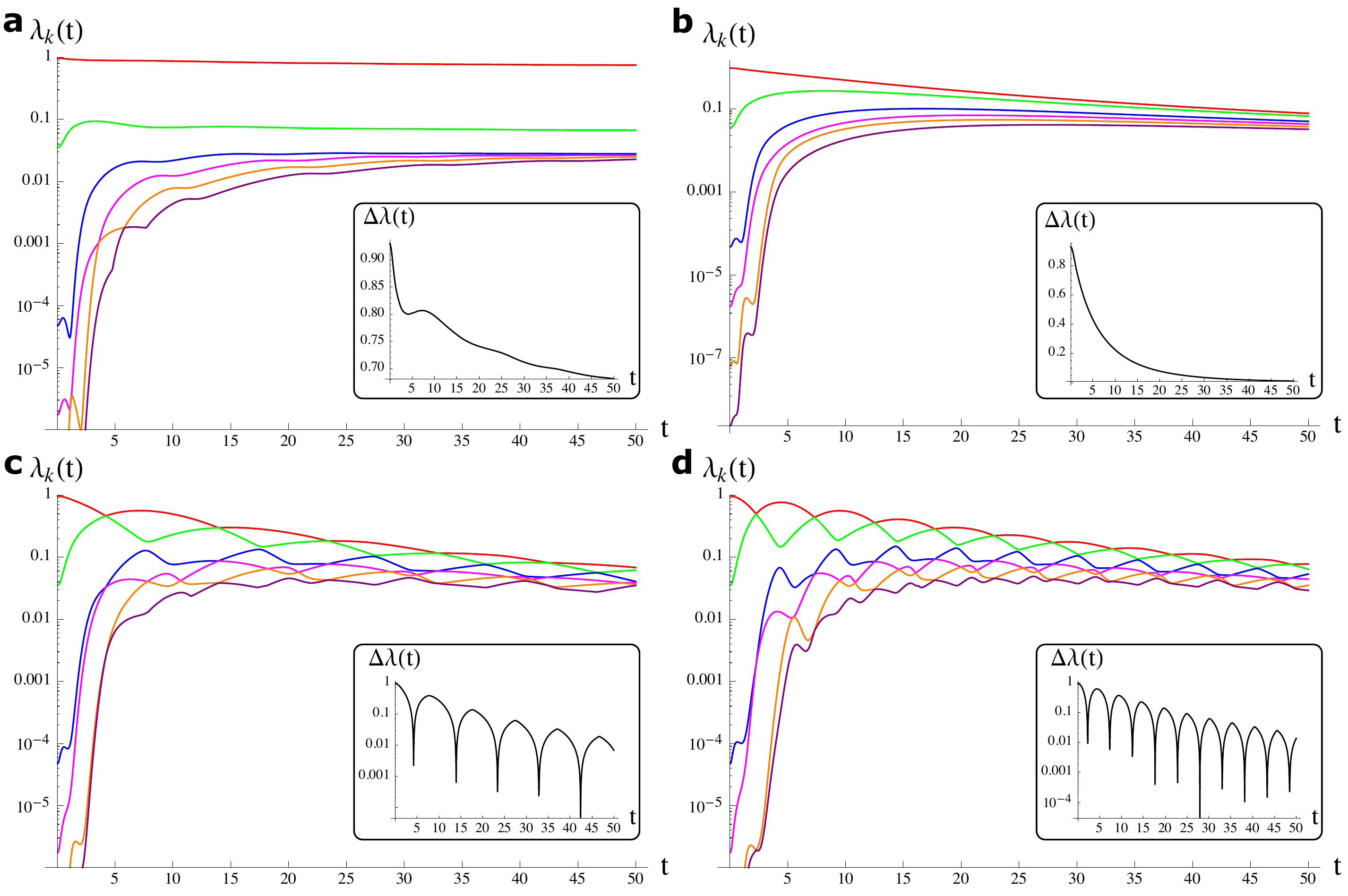}
\caption{Dynamics of entanglement spectrum and Schmidt gap (inset) for the quenches $h_0=1.5$ to $h_1=1.2$ (\textsf{\textbf{a}}), $1.0$ (\textsf{\textbf{b}}), $0.8$ (\textsf{\textbf{c}}), $0.5$ (\textsf{\textbf{d}}) as a function of time.}
\label{ES_15}
\end{figure}
The first quenches we consider are from the paramagnetic to the ferromagnetic phase, with $h_0=1.5$ and $h_1=1.2,\,1.0,\,0.8,\,0.5$. As show in Fig.~\ref{ES_15} the ES dynamics strongly depends on whether the system is quenched within the same phase (paramagnetic in this case) or in the other. On one hand, when the final Hamiltonian magnetic field $h_1\ge1$ pertains to the paramagnetic phase, the Schmidt gap decreases almost monotonically (insets {\bf{a-b}}).  On the other hand when $h_1<1$, the Schmidt gap oscillates, with a series of zeros (insets {\bf{c-d}}), inside the ferromagnetic phase. In the latter case, the ES exhibits, a series of crossings in time between the eigenvalues $\lambda_k(t)$. We observe that the frequency of the crossings increases lowering $h_1$ and going further deep in the ferromagnetic phase while it vanishes as $h_1\to 1$ at the quantum critical point.
\begin{figure}[h]
\centering
\includegraphics[width=160mm]{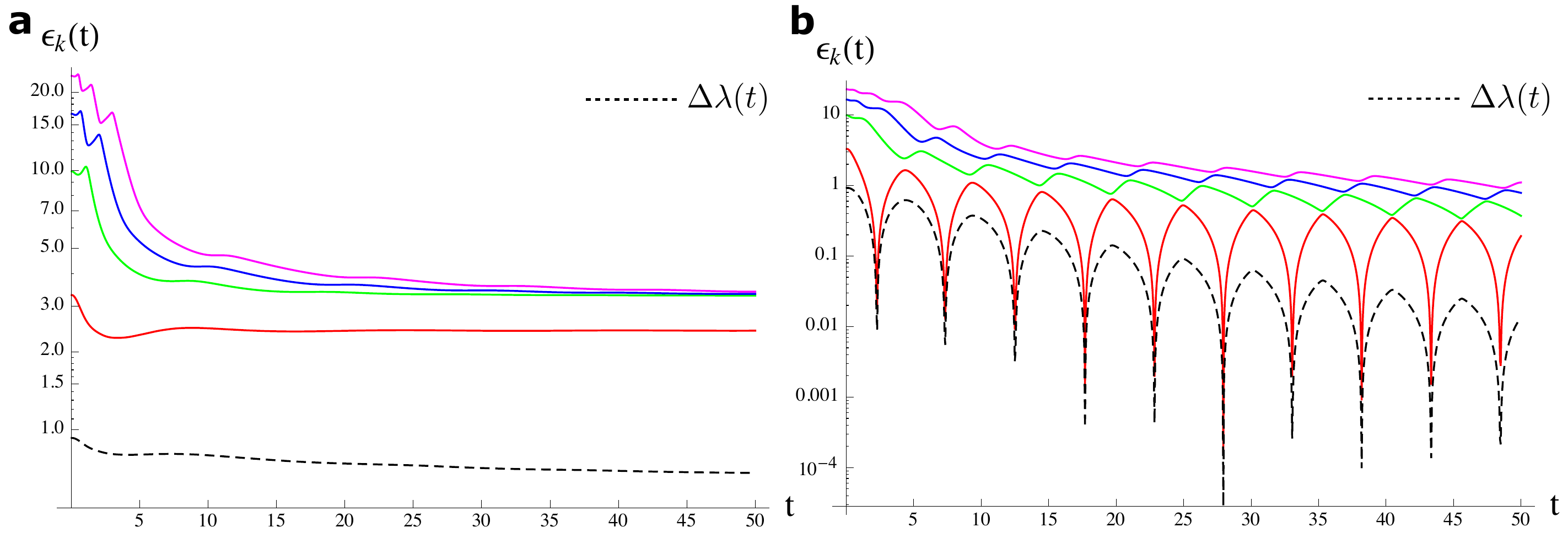}
\caption{Time evolution of the first four eigenenergies $\varepsilon_k$ for the quenches $h_0=1.5$ to $h_1=1.2$ (\textsf{\textbf{a}}) and $0.5$ (\textsf{\textbf{b}}). The black dashed curve in \textsf{\textbf{b}} is the Schmidt gap.}
\label{epsilon_15}
\end{figure}
Such behaviour follows from the single particle eigenvalues $\varepsilon_k$ of $\hat{K}$,  (for a definition see Eq \ref{eq:epsilon} in \ref{sec:ff}) since the smallest one (which is the most important due to the exponential form of the reduced density matrix) goes periodically to zero, as plotted in Fig.~\ref{epsilon_15}. As we see, the Schmidt gap (black dashed curve) matches perfectly the oscillating behaviour of the lowest eigenvalue $\varepsilon_1$.

\begin{figure}[b]
\centering
\includegraphics[width=100mm]{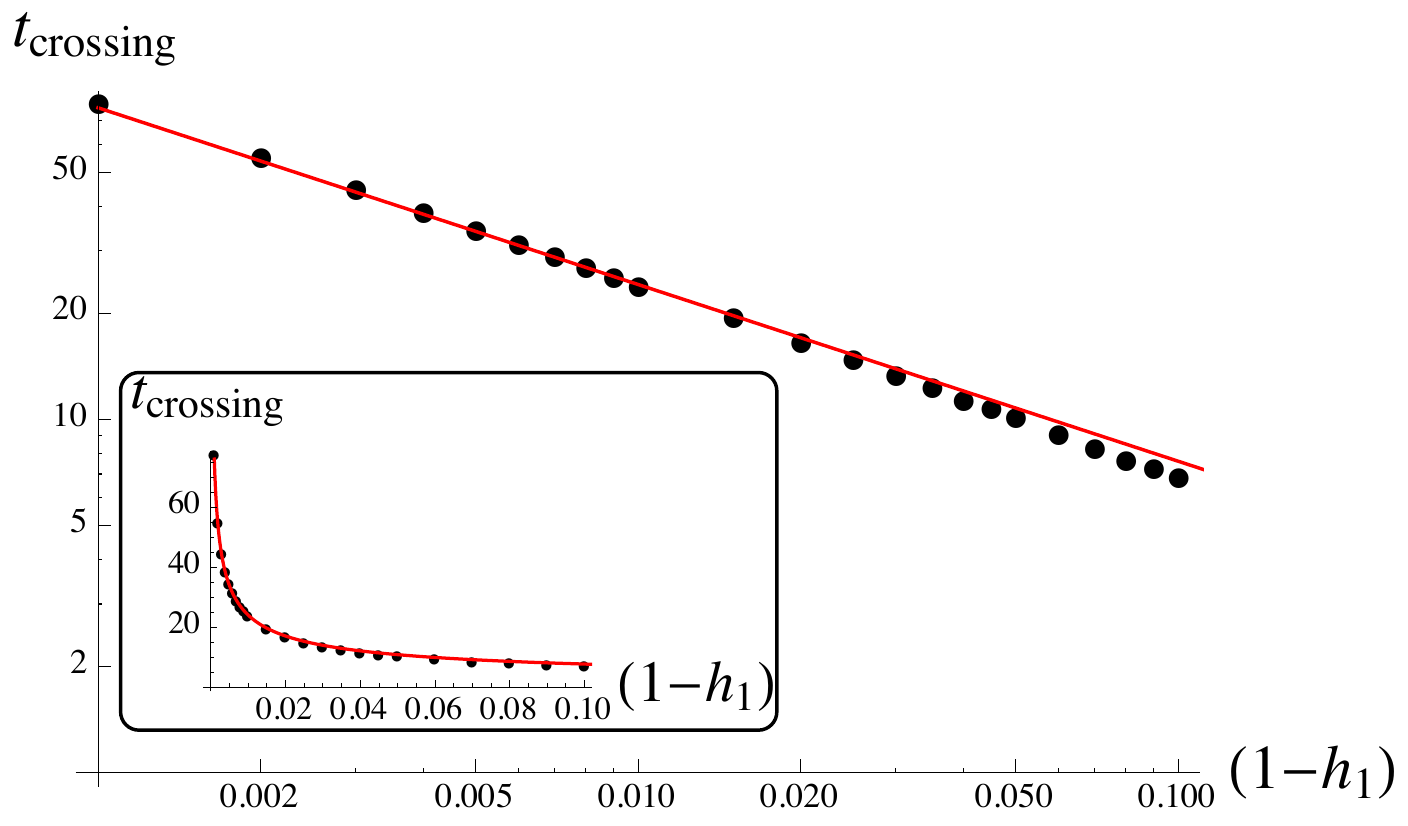}
\caption{Crossing time $t_{\rm crossing}$ at which the first zero of the Schmidt gap occurs as a function of $(1-h_1)$ (dots) and the function $\tilde{t}(h_1)$ (see Eq.~\eqref{eq:tcr1}, solid line) in logarithmic scale. In the inset the crossing time in linear scale. We set $h_0=1.5$.}
\label{t_crossing}
\end{figure}

To better understand the nature of these crossings we can look at the time $t_{\rm crossing}$ when the first crossing between the two largest eigenvalues occurs (i.e. the first zero of the Schmidt gap), as a function of the final Hamiltonian parameter $h_1$. As expected the crossing time diverges algebraically as the post-quench regime gets closer to the quantum critical point. The result is shown in Fig.~\ref{t_crossing} where the numerical data for the crossing times are plotted against $1-h_1$ and compared to the simple scaling function:
\begin{equation}
\label{eq:tcr1}
t_{\rm crossing}(h_1)=\frac{1}{\sqrt{1-h_1}}.
\end{equation}
 Notice that one would be tempted to associate a critical exponent to the power $1/2$ that we find. However this is not the case because as we are quenching the system across the phase transition the evolved quantum state of the system is quite far from equilibrium and thus a description in terms of quasi-equilibrium critical exponents should not be adequate.

We would like to highlight a very close resemblance of the results for the Schmidt gap and those for the spontaneous magnetisation $\langle\hat\sigma^x(t)\rangle$ for a quench across the transition $h_0\to h_1$ \cite{Heyl,Calabrese2012b}. These authors predict and observe numerically that the crossing times go as:
\begin{equation}
t_n^* = t^*\left (n+\frac 12\right), \quad n=0,1,2,\dots
\end{equation}
with $t^*=\pi/E_{k^*}(h_1)$ where $\cos k^*=(1+h_0 h_1)/(h_0+h_1)$ and  
$E_k(h) = \sqrt{(h-\cos k)^2+\sin^2 k}$ are the single particle energies. Now defining $\delta=1-h_1$ and for $\delta\ll 1$we get 
\begin{equation}
t^*\simeq \frac{\pi}{\sqrt{2 \delta}} +O(\sqrt\delta)
\end{equation}
demonstrating that the first oscillation time $t^*$ for the magnetisation diverges indeed with the same scaling as the crossing time $t_{\rm crossing}(h_1)$ for the Schmidt gap of Eq.~\eqref{eq:tcr1}.
 We emphasise, however, that, although the scaling of the time dependence of the two quantities, Schmidt gap and magnetisation is similar, here we are comparing results for quenches in the two opposite directions.

\begin{figure}[h]
\centering
\includegraphics[width=100mm]{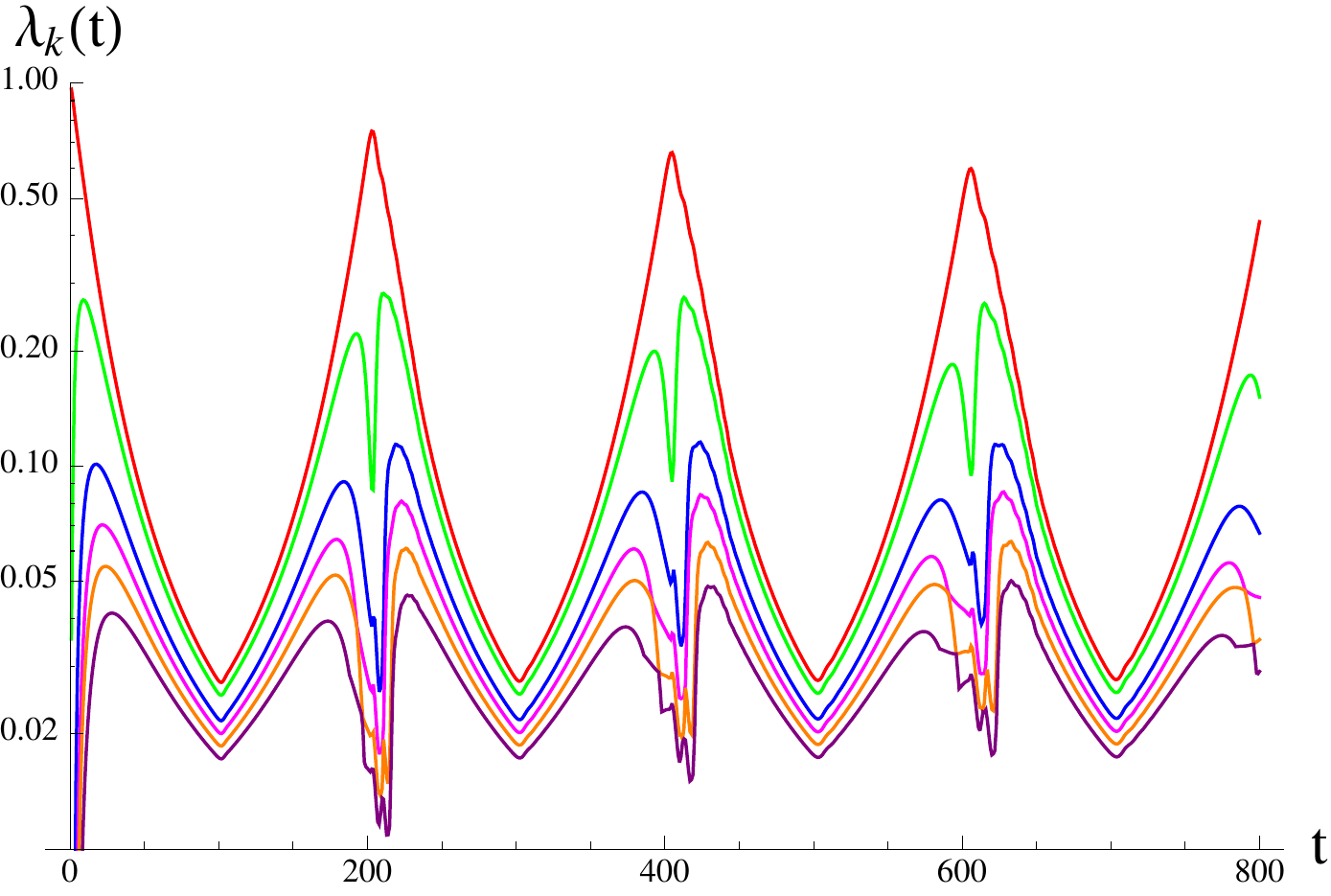}
\caption{Time evolution of the larger six eigenvalues of the reduced density matrix for a chain of $L=200 $ spins, for the quench into the quantum critical point $h:\,1.5\rightarrow 1.0$. The revivals occur approximately every period $T\sim 200$.}
\label{ES_long_time}
\end{figure}

We now turn to the analysis of the effects of the finiteness of the spin chain we consider. For short times we have not observed significant variation with the size of the system for sufficiently long chains. 
However for long times we observe revivals similar to the ones observed for the entanglement entropy. For the quench from $h_0=1.5$ to the quantum critical point $h_1=1$ the results are plotted in Fig.~\ref{ES_long_time} for the six biggest eigenvalues $\lambda_k$: revivals occur at the time $t=L$. Since the initial state is the ground state of the Hamiltonian $\hat{H}_0$, when we quench the Hamiltonian, it will have a higher energy with respect to the new ground state of the Hamiltonian, and will act as a source of quasi-particle excitations along the chain~\cite{Calabrese2005}. The excitations travel along the chain with velocity $v\le1$ (from the dispersion relation) and when they reach the end of the chain they are reflected backward, giving rise to such revivals at time $t=L$. This is the time for a quasi-particle emitted in the middle of the chain to reach the hard wall at the end of the chain, be reflected and come back at the centre. 

\begin{figure}[h]
\centering
\includegraphics[width=16cm]{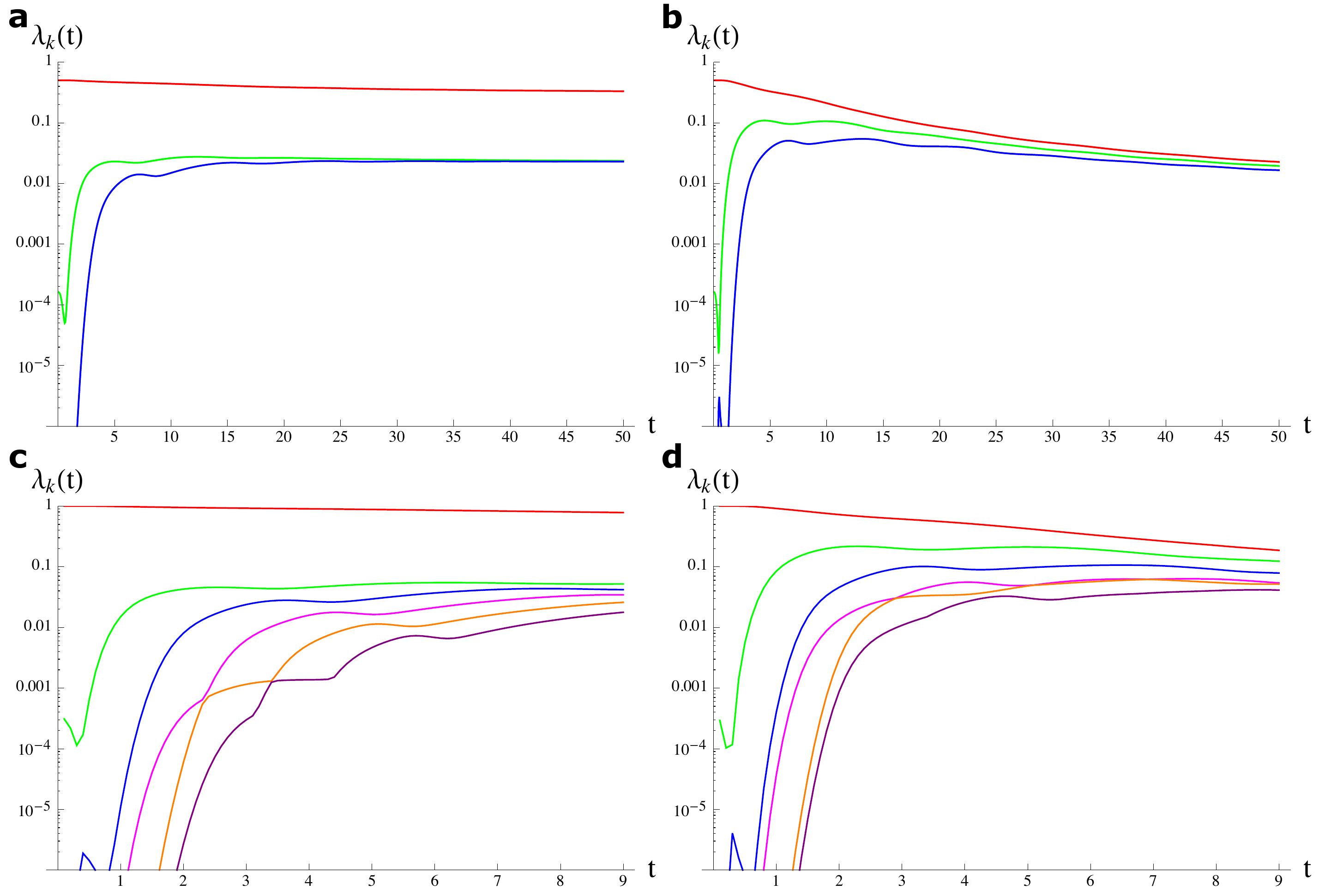}
\caption{ Dynamics of entanglement spectrum for the quenches $h_0=0.5$ to $h_1=0.8$ (\textsf{\textbf{a-c}}), $1.2$ (\textsf{\textbf{b-d}}). Top panels: ES calculated using the exact solution for L=200. Notice that all levels are doubly degenerate. Lower panels: ES calculated using tDMRG for $L=60$.}
\label{quenches_fer}
\end{figure}

From the large time behaviour of the entanglement spectrum we can also
infer some properties of the equilibration ensemble. The Ising model
should equilibrate to a Generalized Gibbs states.  Generalized Gibbs
states are exponential of extended Hamiltonians, that include, apart
from the system Hamiltonian,  also other operators  encoding the
local density of the  conserved quantities.
They thus give rise to extended phase diagrams that depend on several
parameters. The values of the parameters in the generalized
Hamiltonian are fixed by the values of the
conserved quantities in the initial state.
It has been conjectured that these generalized Hamiltonians are responsible
for the appearance of dynamical quantum phase transitions~\cite{Fagotti}.
This kind of transitions emerges when the combination of the
initial state and quench Hamiltonian produce a generalized gapless
Hamiltonian.

Our data in Fig.~\ref{epsilon_15} seem to confirm such conjecture.  At least in this
simple case, when the system is quenched  across a critical point, the
generalized Hamiltonian (that is encoded in the logarithm of the
reduced density matrix for large enough times), becomes gapless (Fig.~\ref{ES_15}
panels b, c, d) while it displays a well defined gap when the system
is quenched inside the same phase Fig.~\ref{ES_15} panel a.

So far we have considered quenches from the paramagnetic phase to the ferromagnetic one and we have seen in the case of a quench connecting two different phases the occurring of crossing between the components of the ES. Such characteristic is surprisingly not found in the opposite case, where the initial state belongs to the ferromagnetic phase. As shown in Fig.~\ref{quenches_fer} in both cases the degeneracy of the ground state (due to the $Z_2$ symmetry of the Hamiltonian in the ferromagnetic phase) is not lifted in the dynamics, and no crossing occurs.
We have checked that the absence of crossing is not a consequence of the unbroken symmetry initial state. Using tDMRG and starting with a broken symmetry state with net magnetisation, we computed the ES after similar quenches. The results are reported in the lower panels of Fig.~\ref{quenches_fer}. As in the case with no symmetry breaking, the two largest eigenvalues never cross (although some crossing are observed for lower eigenvalues).

Despite the lack of crossings, even for the quench from the
paramagnetic phase to the ferromagnetic phase, we appreciate that the
generalized Hamiltonian describing the long time Generalized Gibbs
State is gapped for quenches in the same phase and gapless for
quenches across the phase transition.

\begin{figure}[h]
\centering
\includegraphics[width=7cm]{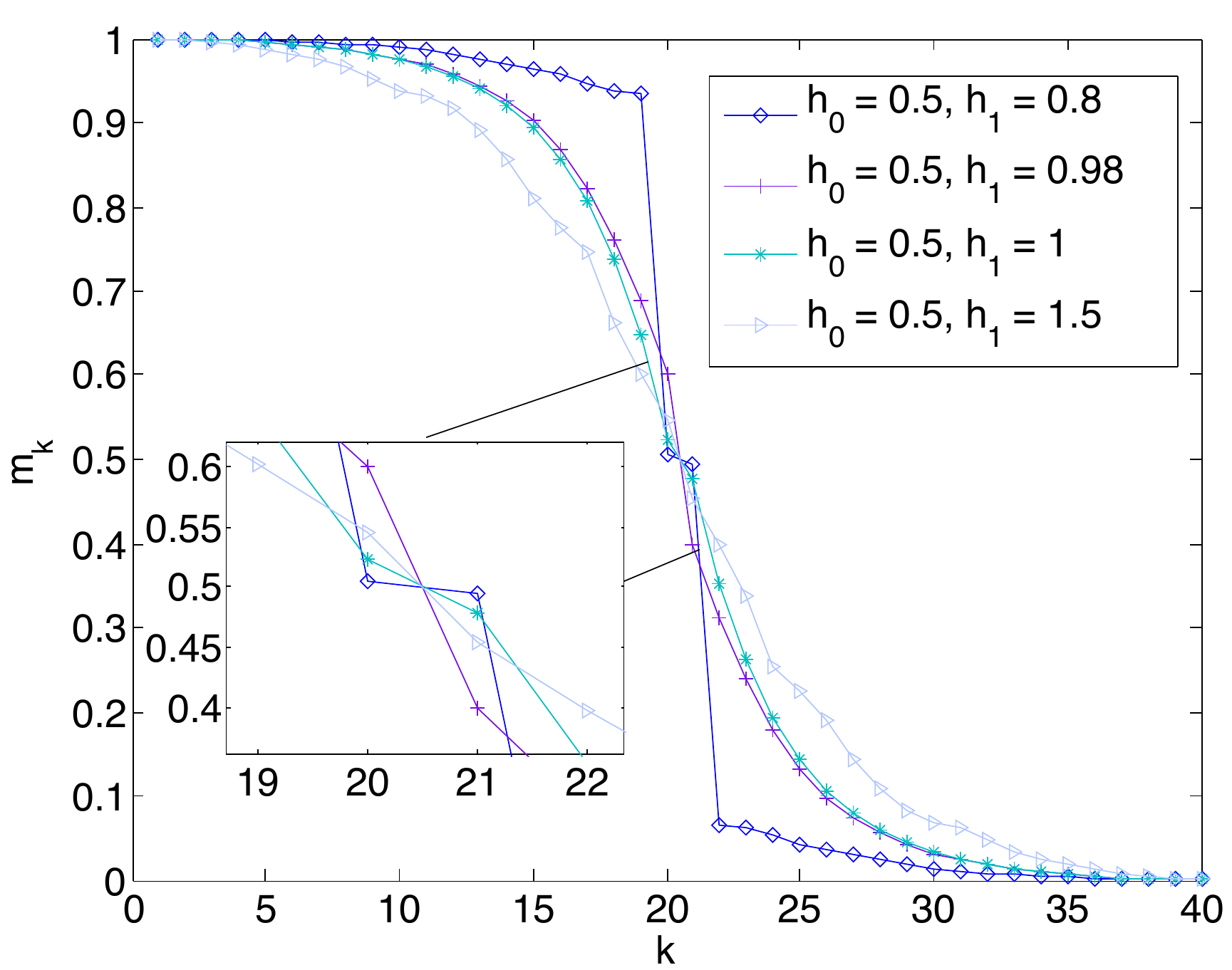}
\includegraphics[width=7cm]{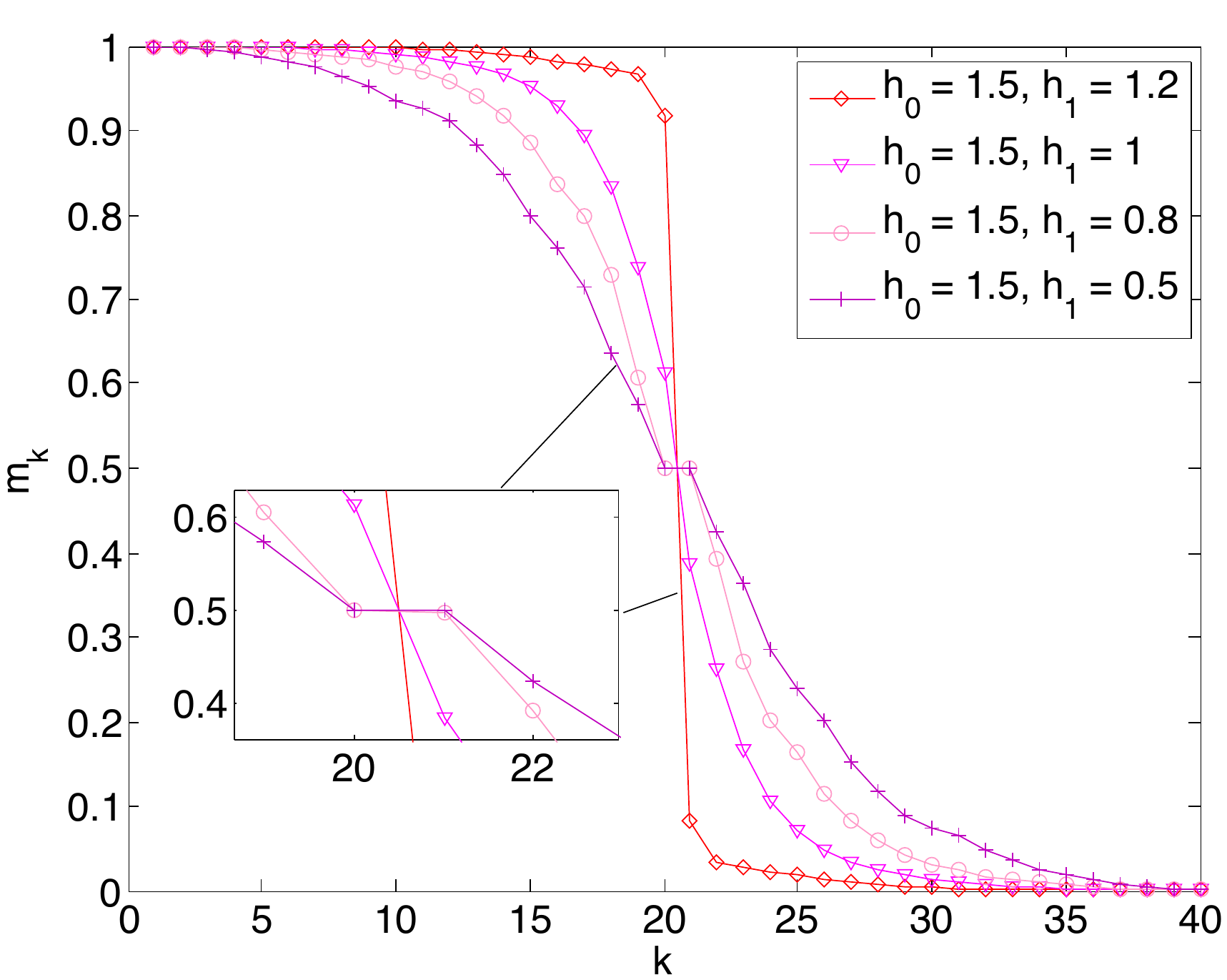}
\caption{Plot of the values $m_k$ as a function of $k$ for  a block of $l=20$ spins in a chain of $N=200$ spins in the diagonal ensembles corresponding to the specific quench protocols. From these plots one can infer the presence or absence of both gap and degeneracies in the entanglement spectrum of the block at very long times after the quench (see \ref{sec:ff}). {\bf Left panel:} Quenches starting from the ferromagnetic phase. When the quench ends in the same phase, the entanglement spectrum is doubly degenerate (due to the presence of two degenerate $m_k$ at $k=L/10$ and $k=L/10+1$) and gapped. On the other hand, once the quench crosses the phase transition it becomes gapless. In the inset we show the relevant modes.  {\bf Right panel:} When the quench starts in the paramagnetic phase and ends in the same phase we observe a gapped entanglement spectrum (there is a gap in the $m_k$ close to $k=L/10$) while for quenches across the phase transition the entanglement spectrum becomes gapless. The inset zooms on the two relevant modes.}
\label{fig:ES_longtimes}
\end{figure}

In order to confirm that the entanglement spectrum becomes gapless once
one crosses a phase transition, we have computed the  entanglement
spectrum of a small block that should be described at large times by the corresponding  diagonal ensemble (DE) \cite{fagotti2013}\footnote{We have checked that the same qualitative  behavior is also observed in the half chain entanglement spectrum obtained from the same DE but in that case the DE is not expected to reproduce the correct quantitative behaviour of half  spin chain  since it should correctly capture the local physics only.}.
Let us consider thus a small block, and compute  the entanglement spectrum of a block of $L/10$ spins
in a chain of $L$ spins.
This can be characterized by plotting the corresponding $L/5$ eigenvalues $m_k$ of the covariance matrix defined in Eq.~\eqref{eq:covariance}, $k=1,\dots, L/5$ versus $k$. The resulting plots are shown in Fig.~\ref{fig:ES_longtimes}. Any element of the entanglement spectrum is obtained by multiplying together $L/10$ of those $m_k$.

The presence or absence of a gap between the first two eigenvalues of
the reduced density matrix  is  related to the presence or
absence of a gap in the plots of Fig.~\ref{fig:ES_longtimes} close to $L/10$. A gap there means
that the two largest eigenvalues are different (since they are obtained
by choosing in the product one or the other, $m_{L/10}$ or
$m_{L/10+1}$).

Still, first and second eigenvalues could be degenerate and a gap might exist
between the second and the third. This is the typical case for $Z_2$
invariant theories where the eigenvalues of the reduced density matrix
come in degenerate pairs.
In this case the gap is now between $m_{L/10-1}$ or $m_{L/10+2}$.

Both cases are observed for quenches in the same phase. Gapped spectra
are obtained for quenches in the paramagnetic phase while gapped
degenerate spectra are obtained for quenches in the ferromagnetic
phase as reported in the two panels Fig. \ref{fig:ES_longtimes}.
As soon as one studies quenches between two different phases, we
observe that the entanglement spectrum becomes really gapless as shown
by both panels of Fig.   \ref{fig:ES_longtimes}.

\subsection{Results for the Ising model: adiabatic quench}
\begin{figure}[h]
\centering
\includegraphics[width=120mm]{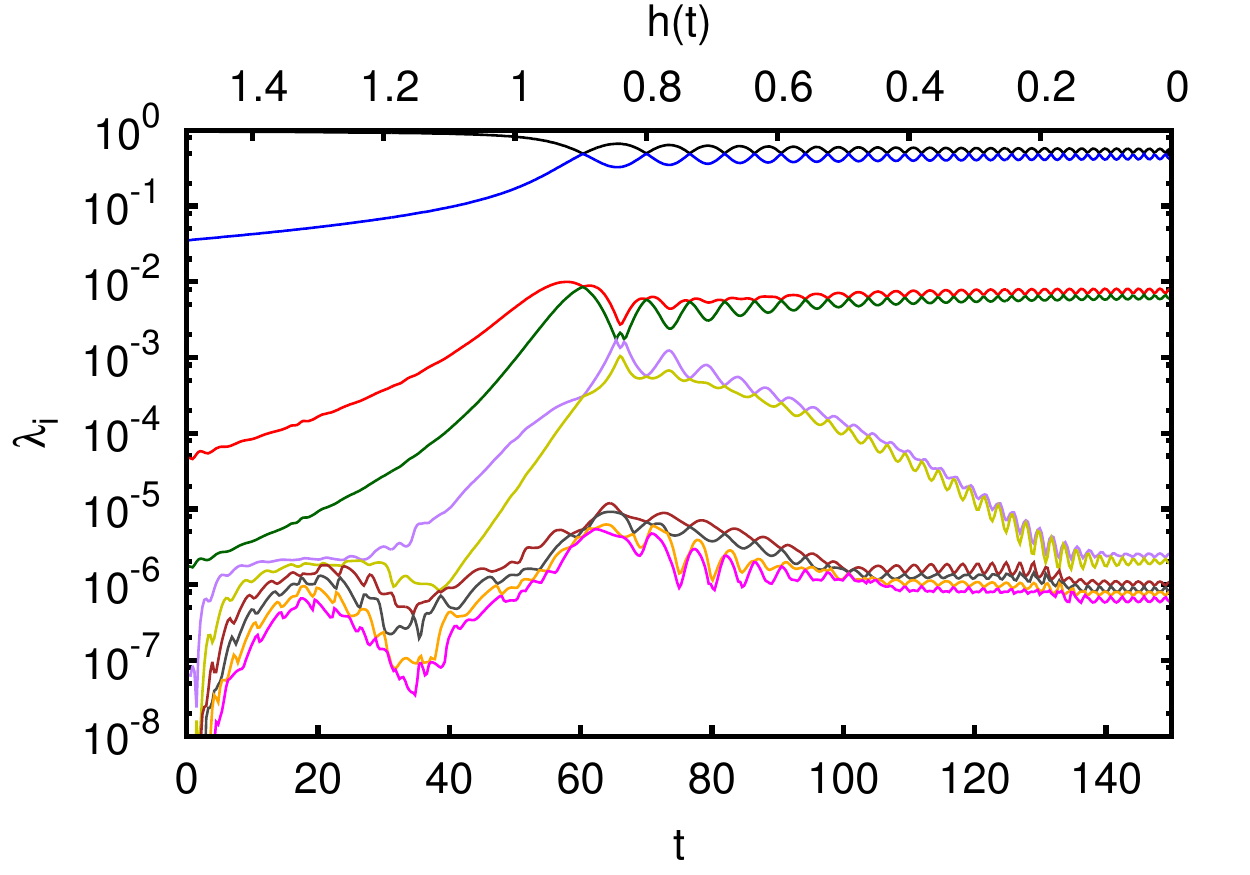}
\caption{Entanglement spectrum dynamics (first 10 eigenvalues) for a quench with initial magnetic field $h_0=1.5$ and rate $\gamma=10^{-2}$ for $L=64$ versus time. The final magnetic field is $h_1=0$. The top horizontal axis show the instantaneous value of the field $h(t)$ at time $t$.}
\label{fig:ES_adiabatic}
\end{figure}

Here we move to linear quasi adiabatic quenches in which the magnetic field is decreased linearly from $h_0$ to $h_1$ as in Eq.~\eqref{eq:htgamma}.
Many previous works have concentrated on the dynamics of entanglement entropy and its relation to the Kibble-Zurek mechanism \cite{Cincio2007,Caneva2008,PollmannPRE,Majumdar2010}.
Our results of the ES dynamics for a quench from the paramagnetic phase $h_0=1.5$ to the ferromagnetic phase $h_1=0$ with rate $\gamma=10^{-2}$ are shown in Fig.~\ref{fig:ES_adiabatic}.

The system is initially in the ground state of $\hat H_0$ thus it is very close to a product state with all the spins pointing in the $z$ direction. Thus initially $\lambda_1\approx 1$ while all the other eigenvalues are negligible. As the magnetic field decreases the first eigenvalue $\lambda_1$ decreases while all the others increase steadily until the instantaneous field $h(t)$ passes the critical point. After this time the eigenvalues start oscillating and crossing (see Fig.\ref{fig:ES_adiabatic}) in an analogous fashion to the instantaneous quenches. Eventually for long times and far below the critical point, the eigenvalues seem to tend to an asymptotic limit with residual oscillations as expected invoking the assumption of the generalised Gibbs ensemble. The eigenvalues distribution gives correctly the value for the von Neumann entropy predicted for slow quench (not shown).

We can see from Fig.~\ref{fig:ES_adiabatic} that the first crossing does not occur for $h(t)=1$ but slightly later in time. The reason for this is twofold: first, close to the phase transition the reaction times of the system compete with the rate of change $\gamma$ of the magnetic field, so there is a finite delay; second, for a finite system, the Schmidt gap of the ground state of the Ising model at the critical point is not zero but approaches zero with an algebraic decay\footnote{With logarithmic corrections.} ruled by the ratio of critical exponents $\beta/\nu = 1/8$ as found in \cite{DeChiara2012,Lepori2013}.

\begin{figure}[h]
\centering
\includegraphics[width=120mm]{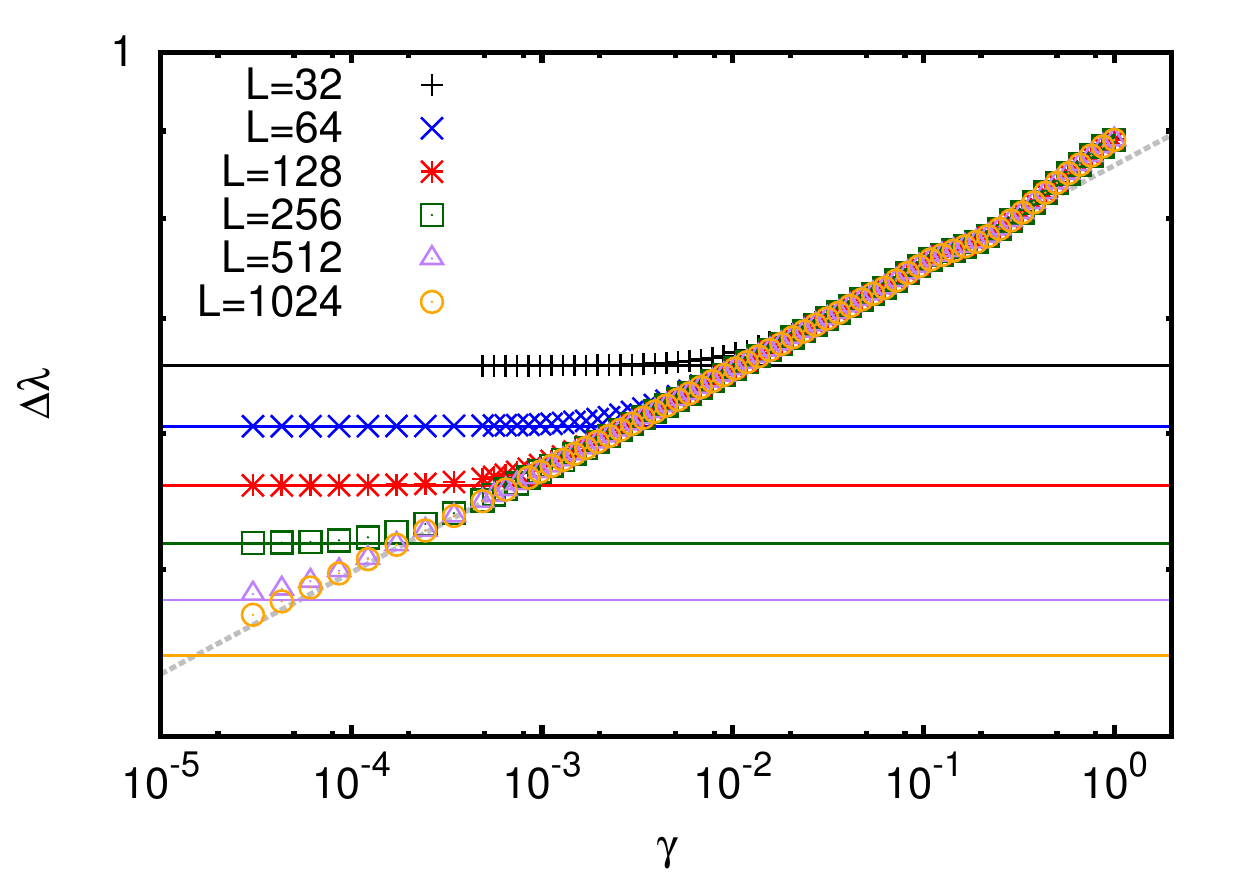}
\caption{Schmidt gap computed at the critical point $h(t)=1$ for different values of $\gamma$ and different chain lengths $L$. The horizontal lines represent the asymptotic $\Delta\lambda$ at the critical point. This is computed from the ground state of the Ising Hamiltonian $\hat H(h=1)$ at the critical point. The system starts always at $h_0=1.5$. The gray dashed line is a power law fit of $L=256$ with power $0.064$.}
\label{fig:DL_gamma}
\end{figure}

These two effects combined together give rise to an interesting behaviour of the Schmidt gap $\Delta\lambda$ when the instantaneous field $h(t)=1$ is at the critical point. The results for $\Delta\lambda$ at the critical point versus the rate of change $\gamma$ of the magnetic field is shown in Fig.~\ref{fig:DL_gamma} for different lengths. For very large values of $\gamma$ the value of $\Delta\lambda$ remains blocked at its initial value. For very small values of $\gamma$ (and not too big sizes), the quench becomes adiabatic and the state remains in the ground state of the instantaneous Hamiltonian. The Schmidt gap at the critical point is therefore very close to the value obtained from the ground state of the Ising Hamiltonian $\hat H(h=1)$ at the phase transition. For intermediate values of $\gamma$, the Schmidt gap $\Delta\lambda$ decreases algebraically $\Delta\lambda \approx \gamma^a$ with a small power $a=0.064$. 

\section{Instantaneous quenches in the XXZ model}
\label{sec:XXZ}

\begin{figure}[h]
\centering
\includegraphics[width=60mm]{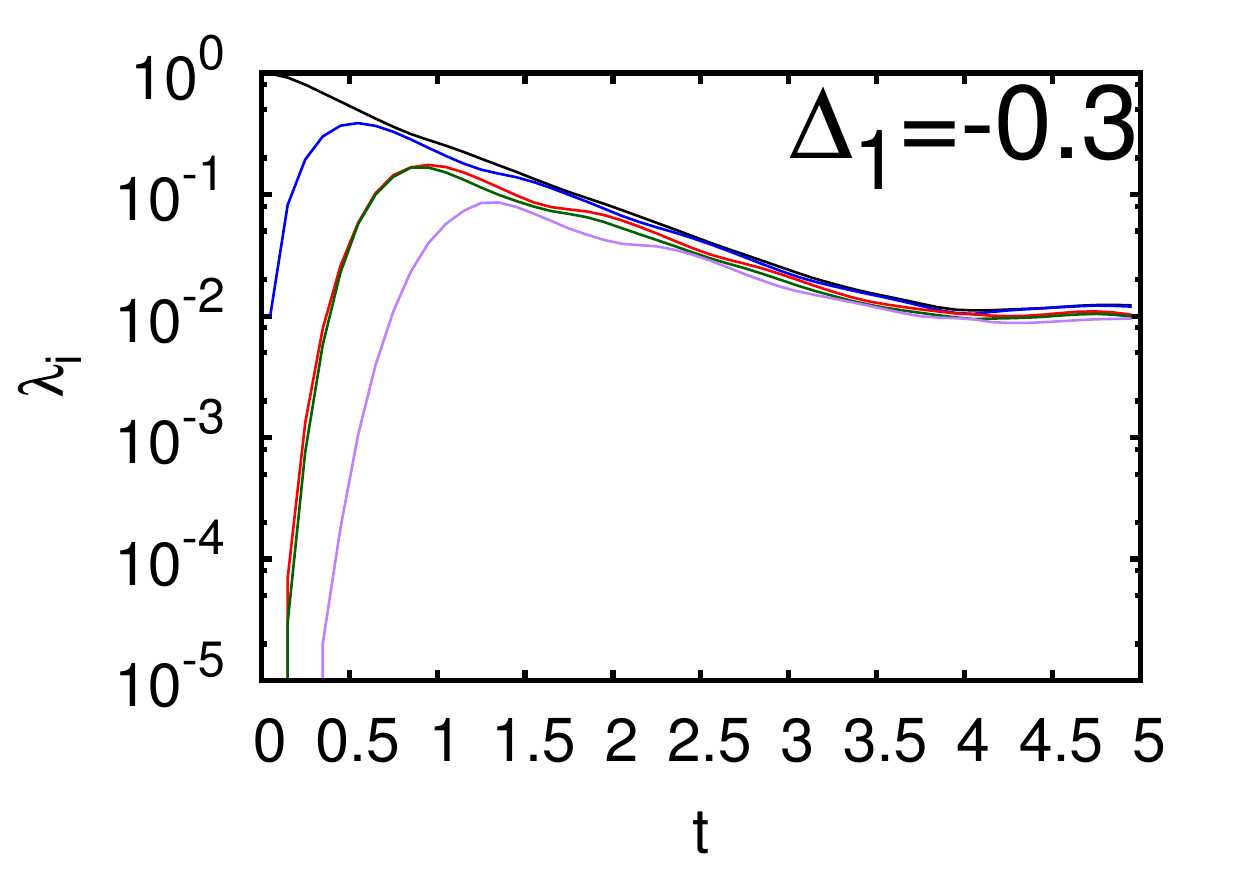}
\includegraphics[width=60mm]{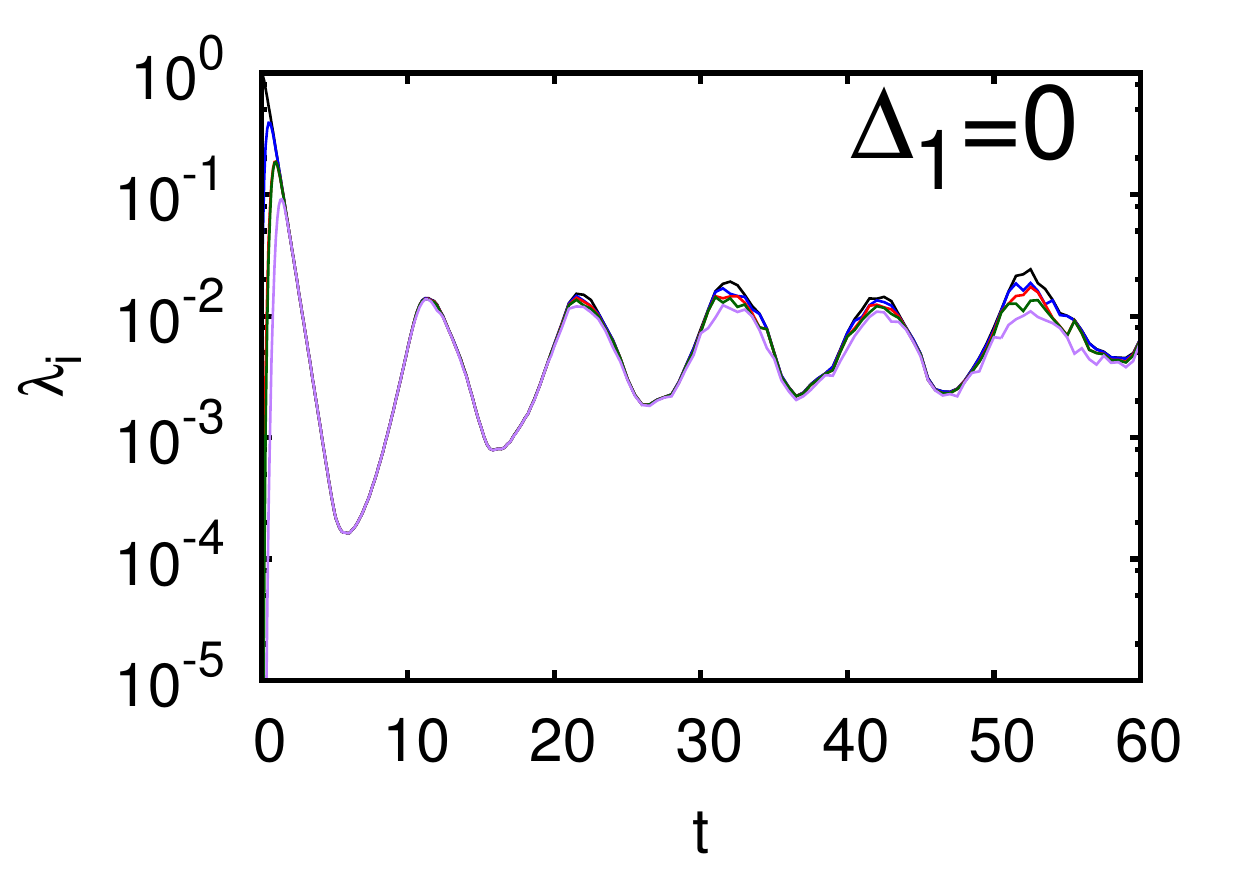}
\\
\includegraphics[width=60mm]{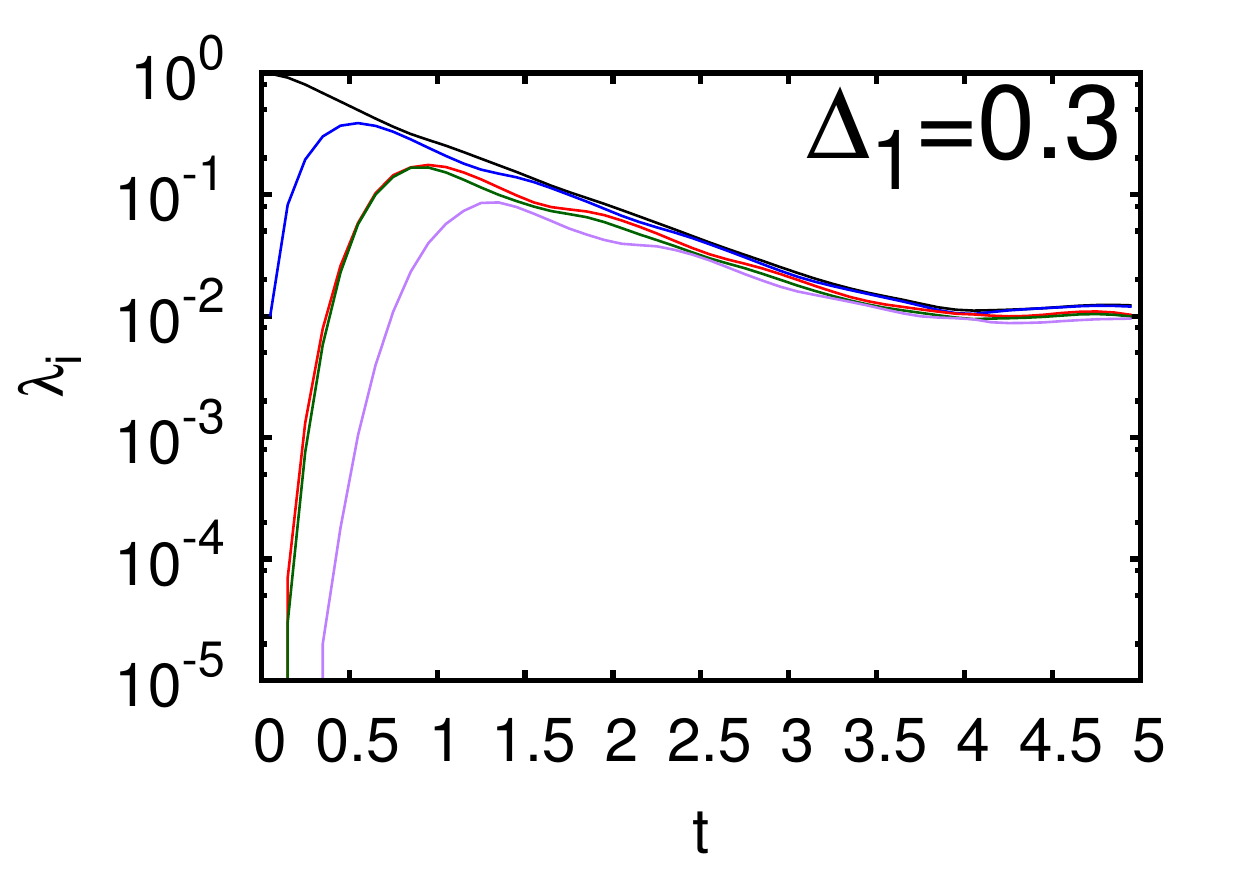}
\includegraphics[width=60mm]{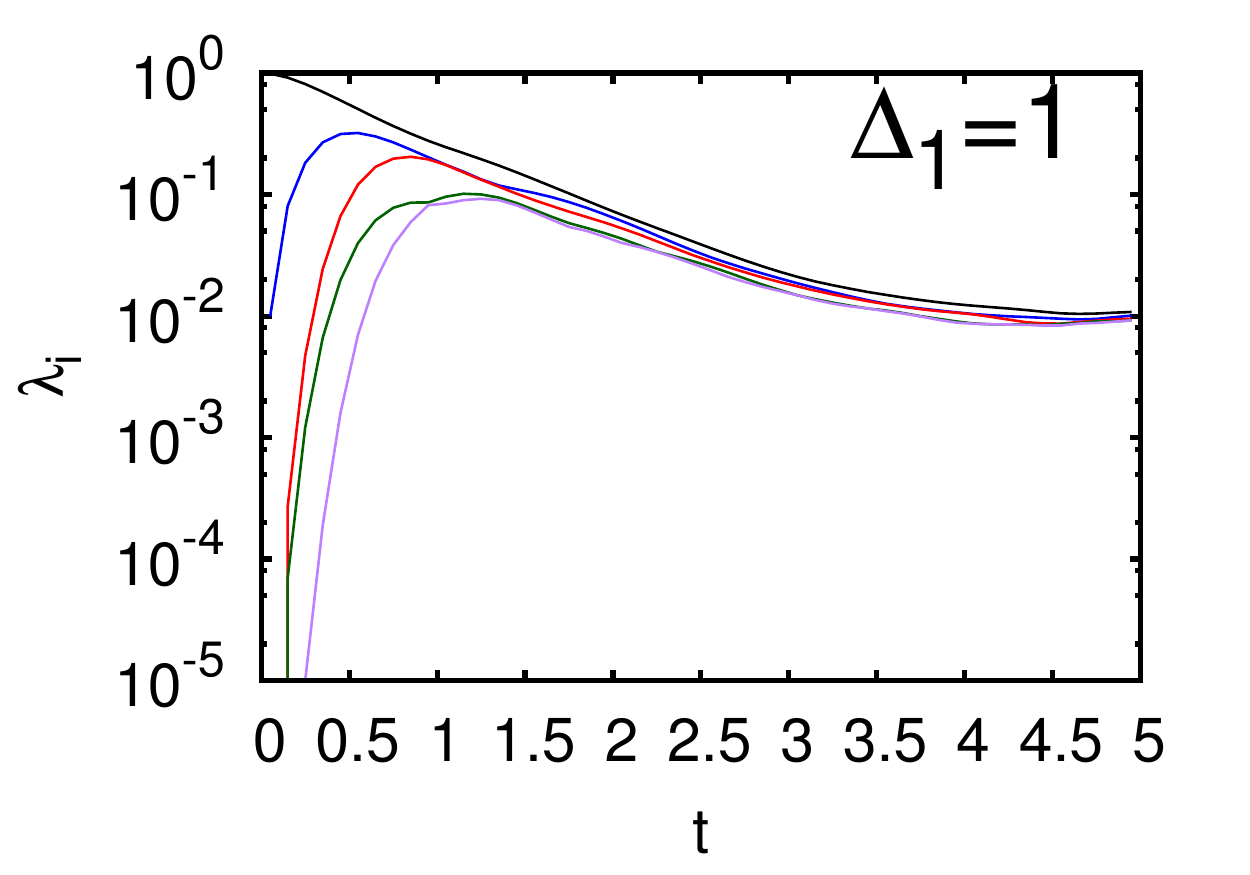}
\caption{Entanglement spectrum dynamics for the XXZ starting from a perfect N\'eel antiferromagnetic state and evolved with the Hamiltonian with $\Delta_1=-0.3,0,0.3,1$ (XY-critical phase). The first 5 eigenvalues computed for $L=40$ are plotted against time.}
\label{fig:XXZ}
\end{figure}

In this section we turn our attention to the XXZ model. The basic properties of the phase diagram have been described in Sec.~\ref{sec:prel}. In contrast to the Ising model that is characterised by an isolated critical point, it exhibits an entire critical region $|\Delta|<1$.  The dynamics of the von Neumann entropy of the evolved state after a quench in the anisotropy from a value $\Delta_0$ to a value $\Delta_1$ has been first studied in Ref.~\cite{DeChiara2006}. Here we extend that analysis to the ES for quenches starting in a perfect N\'eel antiferromagnetic state. To calculate the ES, we distinguish two cases. For $\Delta_1 = 0$ the final Hamiltonian is the XX one and can be mapped into free fermions as for the Ising model and the ES can be computed analogously (see \cite{DeChiara2006}). For $\Delta_1 \neq 0$ we employ the tDMRG method \cite{dmrg} to compute the time evolution of the initial state with Hamiltonian  $\hat{H}(\Delta_1)$. In this section we fix $L=40$. We consider values of $\Delta_1$ in or at the boundary of the critical region. The results are shown in Fig.~\ref{fig:XXZ}. As for the Ising model the first eigenvalue starts from approximately $1$ and decreases, while all other lower eigenvalues increase. This homogenisation of the eigenvalues gives rise to the linear increase of the von Neumann entropy. We do not observe dramatic crossings as for the Ising model. Instead the eigenvalues, after a fast transient, tend to decay together until a saturation. 
In fact, around $t^*\approx 4$, also the von Neumann entropy saturates because of the finiteness of the chain size. 
Moreover for $\Delta_1=0$ the eigenvalues become degenerate and decay approximately exponentially with time. 
 For long times, it is possible to show that the covariance matrix $M$ becomes a multiple of the identity matrix leading to a maximally mixed reduced density matrix after any bipartition.

In the case of the XXZ model, our data suggest the appearance of a
gapless generalized Hamiltonian for quenches from the anti
ferromagnetic phase to the boundary of the critical phase ($\Delta =1$)
and in the critical phase ($\Delta >-1$).  This again suggests
the appearance of a dynamical quantum phase transition for such
quenches.
\begin{figure}[h]
\centering
\includegraphics[width=60mm]{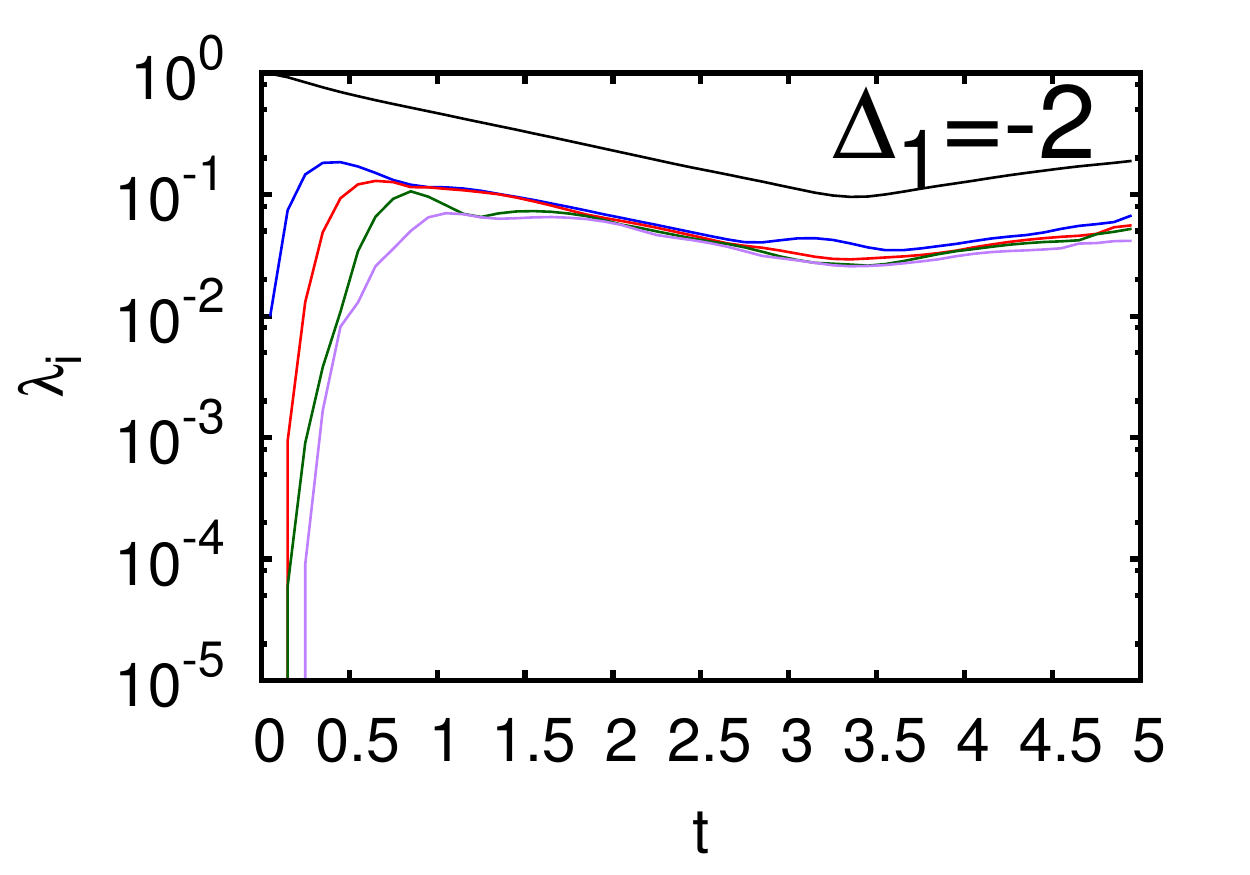}
\includegraphics[width=60mm]{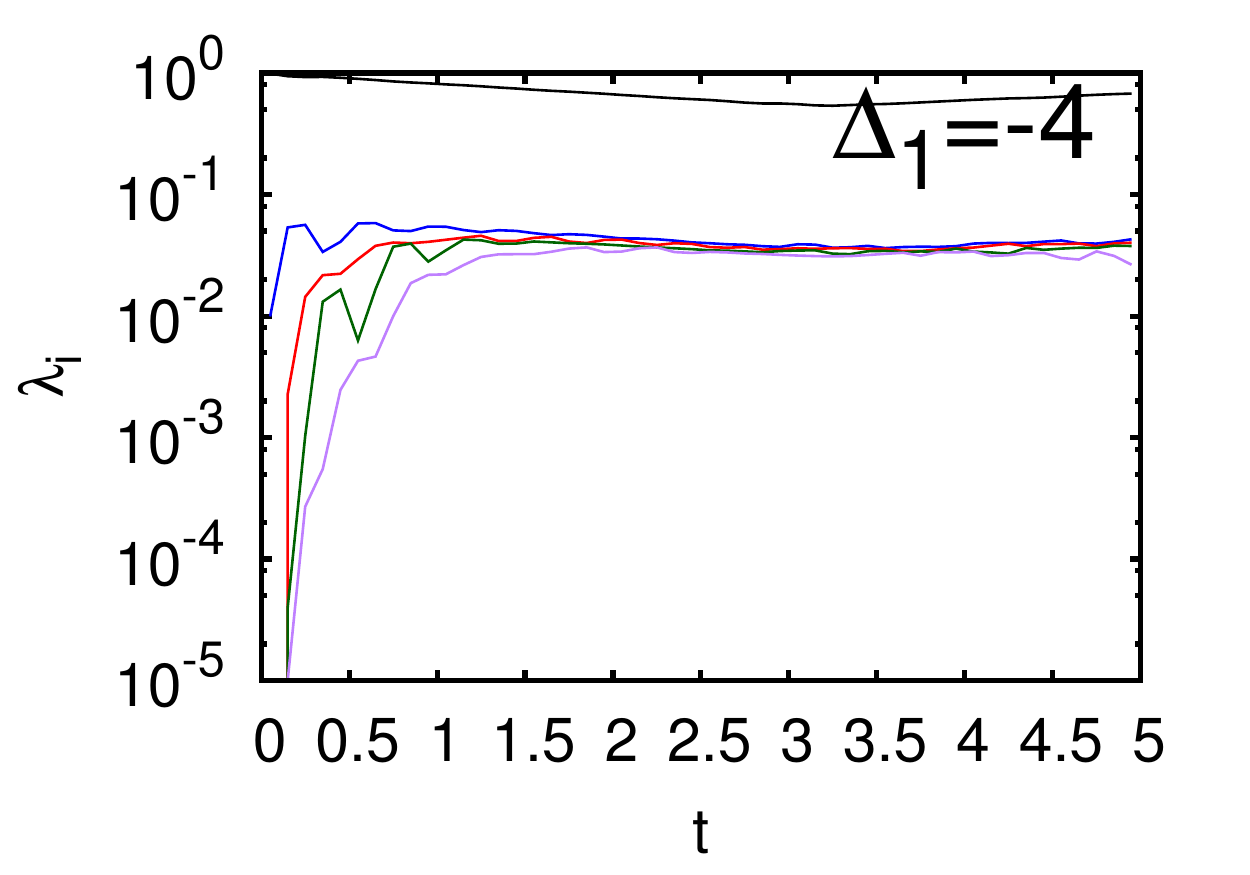}
\caption{Entanglement spectrum dynamics for the XXZ starting from a perfect N\'eel antiferromagnetic state and evolved with the Hamiltonian with $\Delta_1=-2,-4$ (ferromagnetic phase). The first 5 eigenvalues computed for $L=40$ are plotted against time.}
\label{fig:XXZ2}
\end{figure}
Quenching in the ferromagnetic phase ($\Delta<-1$), on the other hand  seems to
produce a gapped generalized Hamiltonian (Fig.~\ref{fig:XXZ2}). This suggests the absence of
a dynamical phase transition despite our quench crossing two static
quantum phase transitions (one at $\Delta =1$ and another one at $\Delta
=-1$).

\section{Conclusions}
\label{sec:conc}
In summary, we have presented an analysis of the entanglement spectrum dynamics in 1D spin chains. We have shown results for instantaneous quenches for the Ising and XXZ models. For the Ising model, we have shown how the dynamics of the Schmidt gap resembles very closely that of the spontaneous magnetisation, thus establishing a further connection with order parameters. Moreover, we have studied quasi-adiabatic quenches, in which the Schmidt gap scales algebraically with the rate of change of the magnetic field. 

Our results demonstrate once more the ability of the ES to reveal interesting features of many-body system close to critical points.
By analyzing the presence/absence of a gap in the long time evolution
of the entanglement spectrum we have confirmed that in the simplest
models dynamical quantum phase transitions seem to arise when
quenching across a quantum critical point. More interestingly, in the
XXZ model, we have observed the absence of a dynamical phase
transition in a quench protocol across several quantum phase
transitions. This result  seems to confirm the conjecture of Ref.~\cite{Fagotti}
that identifies the generalized Hamiltonian, rather than the system
Hamiltonian as the one inducing dynamical phase transitions.

Many open questions remain that go beyond the scope of this work. For example it is natural to ask whether the ES dynamics across other Ising-type transitions, like the N\'eel-Haldane transition in spin-1 chains, behave similarly to the calculations showed here.

{\it During the preparation of this manuscript we became aware of a related and independent work on entanglement spectrum dynamics in the Ising model \cite{Elena}.}

\ack
We thank Elena Canovi, Maurizio Fagotti, Rosario Fazio, Maciej Lewenstein, Ingo Peschel and Anna Sanpera for useful discussions. We acknowledge for financial support the John Templeton Foundation (grant ID 43467), EPSRC (EP/L005026/1, EP/K029371/1), and the EU through the Collaborative Project TherMiQ (Grant Agreement 618074), the ERC QUAGATUA and ERC OSYRIS.

\appendix \label{app}

\section{Time evolution after a quench for the Ising model}
\label{sec:ff}
Here we describe the procedure to diagonalise the Ising model and calculate the time evolution of the ES. The eigenenergies and eigenstates of the Hamiltonian in Eq.~(\ref{ising}) can be found exactly using the Jordan-Wigner transformation followed by a Bogoliubov transformation~\cite{LSM,Pfeuty}. The latter requires only the diagonalisation of an $L\times L$ matrix, thus avoiding finding the eigenvalues of the exponentially large Hamiltonian $\hat H$.
The first step is to map the Pauli operators for each site to a set of fermionic operators:
\begin{equation}
\hat c_i = \exp\left[i \pi\sum_{j=1}^{i-1} \hat\sigma_j^z\right]\;\hat\sigma_i^-
\end{equation}
where $\hat\sigma_i^-=\hat\sigma_i^x-i\hat\sigma_i^y$. This mapping transforms the Hamiltonian in a system of free spinless fermions: 
\begin{equation}
\label{quadratic}
\hat{H}=\sum_{i,j=1}^L\,\Bigl[\hat{c}^\dagger_iA_{ij}\hat{c}_j+\hat{c}_i^\dagger B_{ij}\hat{c}^\dagger_j+\mbox{h.c.}\Bigl],
\end{equation}
where the matrices $A$ and $B$ for the Ising model (\ref{ising}) read
\begin{eqnarray}
A_{ij}=-h\delta_{ij}-\frac{1}{2}(\delta_{i+1,j}+\delta_{j,i+1}), \quad i,j=1,\dots,L\\ 
B_{ij}=\frac{1}{2}(\delta_{i+1,j}-\delta_{j,i+1}).
\end{eqnarray}
 Notice that the real matrices $A$ and $B$ are symmetric and anti-symmetric, respectively.

Following \cite{LSM} we can cast the Hamiltonian in diagonal form (neglecting constant terms):
\begin{equation}
\hat{H}=\sum_{k=1}^L\,\Lambda_{kk}\hat{\eta}^\dagger_k\hat{\eta}_k
\end{equation}
after the proper Bogoliubov transformation $B$
\begin{eqnarray}
\left[\begin{array}{c}
\bm{\eta}\\
\bm{\eta}^\dagger
\end{array}\right]=
B
\left[\begin{array}{c}
\bm{c}\\
\bm{c}^\dagger
\end{array}\right]=
\left(\begin{array}{cc}
\Pi&\Gamma\\
\Gamma&\Pi
\end{array}\right)
\left[\begin{array}{c}
\bm{c}\\
\bm{c}^\dagger
\end{array}\right],
\end{eqnarray}
where $\bm{c}=(\hat{c}_1,\dots,\hat{c}_L)^T$. Notice that, as the system is not translational invariant, the label $k$ does not refer to linear momentum. From the commutation relations $\{\bm{\eta},\bm{\eta}\}=\{\bm{\eta}^\dagger,\bm{\eta}^\dagger\}=\bm{0}$ and $\{\bm{\eta}^\dagger,\bm{\eta}\}=\bm{I}$ ($\bm{I}$ being the $L\times L$ identity matrix) we obtain the following constraints on the block matrices $\Pi$ and $\Gamma$:
\begin{equation}
\cases{\Pi\,\Pi^T+\Gamma\,\Gamma^T=\bm{I}\\
\Pi\,\Gamma^T+\Pi^T\,\Gamma=\bm{0}.}
\end{equation}
The single particle energies $\Lambda_{kk}\ge 0$ and the transformation matrices can be directly calculated by the singular-value decomposition of the sum of the matrices $A$ and $B$
\begin{equation}
\Lambda =\psi\,(A+B)\,\phi^T,
\end{equation}
where the column of the matrices $\phi=\Pi+\Gamma$ and $\psi=\Pi-\Gamma$ are the eigenvectors of $\hat{H}$. Due to the quadratic form of the Hamiltonian, the system is fully described by the one-particle Green's function $\mathcal{G}(\bm{c},\bm{c}^\dagger)$
\begin{equation}
\mathcal{G}(\bm{c},\bm{c}^\dagger)=\left\langle\left[\begin{array}{c}
\bm{c}\\
\bm{c}^\dagger
\end{array}\right]  \left[\begin{array}{c}
\bm{c}^\dagger\\
\bm{c}
\end{array}\right]^T   \right\rangle=B^T\mathcal{G}(\bm{\eta},\bm{\eta}^\dagger)B.
\end{equation}

\section{Calculation of the entanglement spectrum dynamics for the Ising model}

Let us first consider the dynamics of the system after an instantaneous quantum quench on the magnetic field. The initial Hamiltonian 
\begin{equation}
\label{eq:H0}
\hat{H}_0=\sum_{k=1}^L\,(\Lambda_0)_{kk}\,\hat{\eta}^\dagger_{k}\hat{\eta}_k.
\end{equation}
at time $t<0$ is abruptly changed to 
\begin{equation}
\hat{H}_1=\sum_{k=1}^L\,(\Lambda_1)_{kk}\,\hat{\xi}^\dagger_k\hat{\xi}_k.
\end{equation}
for $t\ge0$. All the properties of the system, including the ES, can be computed directly from the time-dependent Green's function $\mathcal{G}(\bm{c},\bm{c}^\dagger;t)$. Within the Heisenberg picture, the time evolution is governed by the evolution operator $\hat{U}(t,0)=\mbox{exp}\{-i\sum_{k=1}^L\,(\Lambda_1)_{kk}\,\hat{\xi}^\dagger_k\hat{\xi}_k t\}$. It is then convenient to first transform the operators $(\bm{c},\bm{c}^\dagger)$ into the new set $(\bm{\xi},\bm{\xi}^\dagger)$, for which the time evolution results in a simple phase factor
\begin{equation}
\bm{\xi}(t)=\e^{-i\Lambda_1 t}\,\bm{\xi}\equiv D_t\,\bm{\xi},
\end{equation}
where the diagonal matrix $D_t$ is defined as $(D_t)_{ij}\equiv\delta_{ij}\,\e^{-i(\Lambda_1)_{ij}t}$.

The operators $\bm{\xi}(t)$ can then be mapped back into the set $(\bm{\eta},\bm{\eta}^\dagger)$, where the Green's function 
\begin{equation}
\mathcal{G}(\bm{\eta},\bm{\eta}^\dagger;0)=
\left(\begin{array}{cc}
\bm{I}&\bm{0}\\
\bm{0}&\bm{0}
\end{array}\right)
\end{equation}
 is trivial since the initial state $\ket{\Psi_0}$ is the vacuum of Hamiltonian (\ref{eq:H0}) and therefore $\bm{\eta}\ket{\Psi_0}=0$. If $B_0$ and $B_1$ are respectively the Bogoliubov transformation that maps the operators $(\bm{c},\bm{c}^\dagger)$ into $(\bm{\eta},\bm{\eta}^\dagger)$ and $(\bm{\xi},\bm{\xi}^\dagger)$, such chain of transformations leads to
\begin{equation}
\eqalign{\mathcal{G}(\bm{c},\bm{c}^\dagger;t)&=B_1^T\,\mathcal{G}(\bm{\xi},\bm{\xi}^\dagger;t)\,B_1\\
&=B_1^T \mathcal{D}(t)\,\mathcal{G}(\bm{\xi},\bm{\xi}^\dagger;0)\,\mathcal{D}^\dagger(t)B_1\\
&=B_1^T \mathcal{D}(t)B_1\,\mathcal{G}(\bm{c},\bm{c}^\dagger;0)\,B_1^T \mathcal{D}^\dagger(t)B_1\\
&=B_1^T \mathcal{D}(t)B_1 B_0^T\,\mathcal{G}(\bm{\eta},\bm{\eta}^\dagger)\,B_0 B_1^T \mathcal{D}^\dagger(t)B_1\\},
\end{equation}
where $\mathcal{D}(t)$ is defined as
\begin{equation}
\mathcal{D}(t)=\left(\begin{array}{cc}D_t&0\\0&D^*_t\end{array}\right)
\end{equation}
Defining the following matrices
\begin{equation}
\cases{X(t)=\Pi_1^T D_t\,\Pi_1+\Gamma_1^T D^*_t\,\Gamma_1\\
Y(t)=\Pi_1^T D_t\,\Gamma_1+\Gamma_1^T D^*_t\,\Pi_1\\
U(t)=X(t)\Pi_0^T +Y(t)\Gamma_0^T \\
V(t)=X(t)\Gamma_0^T +Y(t)\Pi_0^T}
\end{equation}
it is easy to show that the time-dependent Green's function is
\begin{eqnarray}
\mathcal{G}(\bm{c},\bm{c}^\dagger;t)&=\left\langle\left(\begin{array}{cc}
\bm{c}(t)\bm{c}^\dagger(t)&\bm{c}(t)\bm{c}(t)\\
\bm{c}^\dagger(t)\bm{c}^\dagger(t)&\bm{c}^\dagger(t)\bm{c}(t)
\end{array}\right)\right\rangle\equiv\left(\begin{array}{cc}
1-C^*(t)&-F^*(t)\\
F(t)&C(t)
\end{array}\right)\\
&=\left(\begin{array}{cc}
U(t)U^\dagger(t)&U(t)V^T(t)\\
V^*(t)U^\dagger(t)&V^*(t)V^T(t)
\end{array}\right),
\end{eqnarray}
where we introduced the correlation matrices
\begin{eqnarray}
\label{Correlations}
C(t)=\langle\bm{c}^\dagger(t)\bm{c}(t)\rangle=V^*(t)V^\dagger(t), \\
\label{Correlation}
F(t)=\langle\bm{c}^\dagger(t)\bm{c}^\dagger(t)\rangle=V^*(t)U^T(t),
\end{eqnarray}
with $C(t),F(t)\in\Omega\oplus\Omega^\perp$. We now divide the full system into subsystem $\Omega$ and its orthogonal complement $\Omega^\perp$. Since we consider a fermionic system, the eigenstates of the total system $\Omega\oplus\Omega^\perp$ are Slater determinants and thus, according to  Wick theorem, the reduced density matrix for the subsystem $\Omega$ has exponential form \cite{Peschel1}
\begin{equation}
\rho_\Omega=\mbox{Tr}_{\Omega^\perp}\ket{\Psi}\bra{\Psi}=Z\,\mbox{exp}(-\hat{K})
\end{equation}
with $\hat{K}$ quadratic in the Fermi operators
\begin{equation}
\hat{K}=\sum_{i,j=1}^L\,\Bigl[\hat{c}^\dagger_i\alpha_{ij}\hat{c}_j+\frac{1}{2}(\hat{c}_i^\dagger \beta_{ij}\hat{c}^\dagger_j+\mbox{h.c.})\Bigl].
\end{equation}
where the coefficients $\alpha_{ij}$ and $\beta_{ij}$ are related to the correlation matrices $C(t)$ and $F(t)$. Following an analogous procedure employed for diagonalising \eqref{quadratic}, the {\it entanglement Hamiltonian} $\hat{K}$ can be brought to its diagonal form
\begin{equation}
\label{eq:epsilon}
\hat{K}=\sum_{k=1}^L\,\varepsilon_{k}\tilde{\eta}^\dagger_k\tilde{\eta}_k
\end{equation} 
with the appropriate transformation $\tilde{B}$. Since the system is completely characterized by the one-particle Green's function, we can derive the time evolution of the ES $\bm{\lambda}(t)$ of $\rho_\Omega(t)$ directly from the projection onto $\Omega$ of the correlation matrices $C$ and $F$
\begin{equation}
\label{eq:projection}
c(t)=P_\Omega\,C(t)\,P_\Omega\qquad\qquad f(t)=P_\Omega\,F(t)\,P_\Omega\qquad c(t),f(t)\in\Omega.
\end{equation}
If the partition is in real space as in our case, then the projection \eqref{eq:projection} consists in selecting the elements in $C(t)$ and $F(t)$ with indexes belonging to the subsystem $\Omega$.

Such correlation functions are obtained directly from Eqs.~\eqref{Correlations}-\eqref{Correlation} but the same results must be found from
\begin{equation}
\label{corr_trace}
c_{ij}(t)=\mbox{Tr}_\Omega[\rho_\Omega\hat{c}_i^\dagger(t)\hat{c}_j(t)]\qquad f_{ij}(t)=\mbox{Tr}_\Omega[\rho_\Omega\hat{c}_i^\dagger(t)\hat{c}_j^\dagger(t)]
\end{equation}
which provides us the way to relate the eigenvalues $\varepsilon_k$ of $\hat{K}$ (and thus the ES) with the Green's functions components \cite{Peschel2,Peschel3}. 
The eigenvalues of $\hat{K}$ are found from the diagonalisation of the $L\times L$ matrix $M$
\begin{equation}
\label{eq:covariance}
M(t)=
\left(\begin{array}{cc}
2c(t)-\bm{I}&2f(t)\\
-2f(t)^*&-(2c(t)^*-\bm{I})
\end{array}\right)
\end{equation}
whose eigenvalues are $m_k(t)=\pm\tanh(\varepsilon_k(t)/2)$. 

It is important to remind here that once the $m_k$ are known the actual entanglement spectrum is obtained by a product of $L$ of them.  The presence or absence of a gap in the entanglement spectrum is then directly related to the presence or absence of degeneracies in the eigenvalues of $M$ as discussed in the main text.
The reduced density matrix is then
\begin{equation}
\rho_\Omega(t)=\frac 1Z\,\exp\left(-\sum_{k}\varepsilon_k(t)\hat{\eta}^\dagger_k(t)\hat{\eta}_k(t)\right) \label{eq:eps}
\end{equation}
whose eigenvalues are simply:
\begin{equation}
\label{eq:lambdai}
\lambda_i = \frac 1Z \exp\left(-\sum_{k=1}^{L/2}\varepsilon_k(t)n^i_k\right)
\end{equation}
where the numbers $n_k^i=0,1$ and the normalisation constant $Z$ is found by imposing $\sum_{i=1}^{N_{max}}\lambda_i=1$ with $N_{max}=2^{L/2}$. As it is practically impossible to calculate the normalisation constant exactly as soon as $L$ becomes large (we use $32\le L\le 512$), we normalise the eigenvalues restricting the sum appearing in \eqref{eq:lambdai} to $1\le k\le k_{max}$. This limits  the calculation to the first $2^{k_{max}}\le N_{max}$ eigenvalues. In our simulations we observe that $k_{max}\le 8$ gives accurate results to machine precision.

For non instantaneous quenches the Hamiltonian $\hat H(t)$ is changing with time. This means that the fermionic operators of the instantaneous Hamiltonian $\hat H(t)$  also change with time. In order to deal with this slightly more complicated case, one can use a time-dependent Bogoliubov transformation as explained for example in Ref.~\cite{Cincio2007} where a set of ordinary differential equations in time must be solved numerically. We avoid this step by using an explicit finite difference expansion of the evolution. In other words, in the Heisenberg picture, we evolve the Green's function $\mathcal{G}(\bm{c},\bm{c}^\dagger;t)$ for a small time interval $dt$ to obtain $\mathcal{G}(\bm{c},\bm{c}^\dagger;t+dt)$ using the same procedure of the instantaneous quenches case assuming the Hamiltonian $\hat H(t+dt)$ to be constant in the interval $[t;t+dt]$. In the numerical simulations we used $dt=10^{-3}$ which is sufficiently small to obtain accurate results.

\end{document}